\documentclass[aps,prb,twocolumn,10pt,aps]{revtex4-1}
\usepackage{graphicx,amsmath,amssymb}
\usepackage[utf8x]{inputenc}
\usepackage{url}
\usepackage[sectionbib]{chapterbib}

\begin{document}

\title{Photon Management in Two-Dimensional Disordered Media}

\author{Kevin Vynck$^1$}
\author{Matteo Burresi$^{1,2}$}
\author{Francesco Riboli$^1$}
\author{Diederik S. Wiersma$^{1,2}$}

\affiliation{1. European Laboratory for Non-linear Spectroscopy (LENS), University of Florence, Via Nello Carrara 1, 50019 Sesto Fiorentino (FI), Italy.}
\affiliation{2. Istituto Nazionale di Ottica (CNR-INO), Largo Fermi 6, 50125 Firenze (FI), Italy.}


\begin{abstract}
Elaborating reliable and versatile strategies for efficient light coupling between free space and thin films is of crucial importance for new technologies in energy efficiency. Nanostructured materials have opened unprecedented opportunities for light management, notably in thin-film solar cells\cite{Polman2012, Mallick2011}. Efficient coherent light trapping has been accomplished through the careful design of plasmonic nanoparticles and gratings~\cite{Atwater2010, Pillai2010}, resonant dielectric particles~\cite{Yao2012, Spinelli2012} and photonic crystals~\cite{Meng2011, Battaglia2011, Mallick2012, Bozzola2012}. Alternative approaches have used randomly-textured surfaces~\cite{Rockstuhl2010,Ferry2011,Sheng2011} as strong light diffusers to benefit from their broadband and wide-angle properties.
Here, we propose a new strategy for photon management in thin films that combines both advantages of an efficient trapping due to coherent optical effects and broadband/wide-angle properties due to disorder. Our approach consists in the excitation of electromagnetic modes formed by multiple light scattering and wave interference in two-dimensional random media. We show, by numerical calculations, that the spectral and angular responses of thin films containing disordered photonic patterns are intimately related to the in-plane light transport process and can be tuned through structural correlations. Our findings, which are applicable to all waves, are particularly suited for improving the absorption efficiency of thin-film solar cells and can provide a novel approach for high-extraction efficiency light-emitting diodes.

\vspace{0.1in}

\textbf{Published in Nature Materials, vol. 11, pp. 1017--1022 (2012). doi:10.1038/nmat3442}

\end{abstract}

\maketitle

Wave transport in disordered systems is a vast research topic, ranging from electrons in semiconductors to light in random dielectrics~\cite{Akkermans2007}, to cold atoms in laser speckles~\cite{Lye2005,Aspect2009}. In disordered optical materials, the multiple scattering of light and the interferences between propagating waves lead to the formation of electromagnetic modes with varying spatial extent and lifetime, depending on scattering strength, structural correlations, and dimensionality of the system~\cite{Sheng2010, Wang2011}. Two-dimensional disordered structures, in particular, have been the playground over the years for the study of complex optical phenomena, including light localization~\cite{Sebbah1993, Sigalas1996, Schwartz2007, Riboli2011} -- two being the marginal dimension below which all waves, in principle, are localized~\cite{Vollhardt1980} -- and random lasing~\cite{Sebbah2002, Noh2011}.

\begin{figure*}
\begin{center}
	\includegraphics[width=0.8\textwidth]{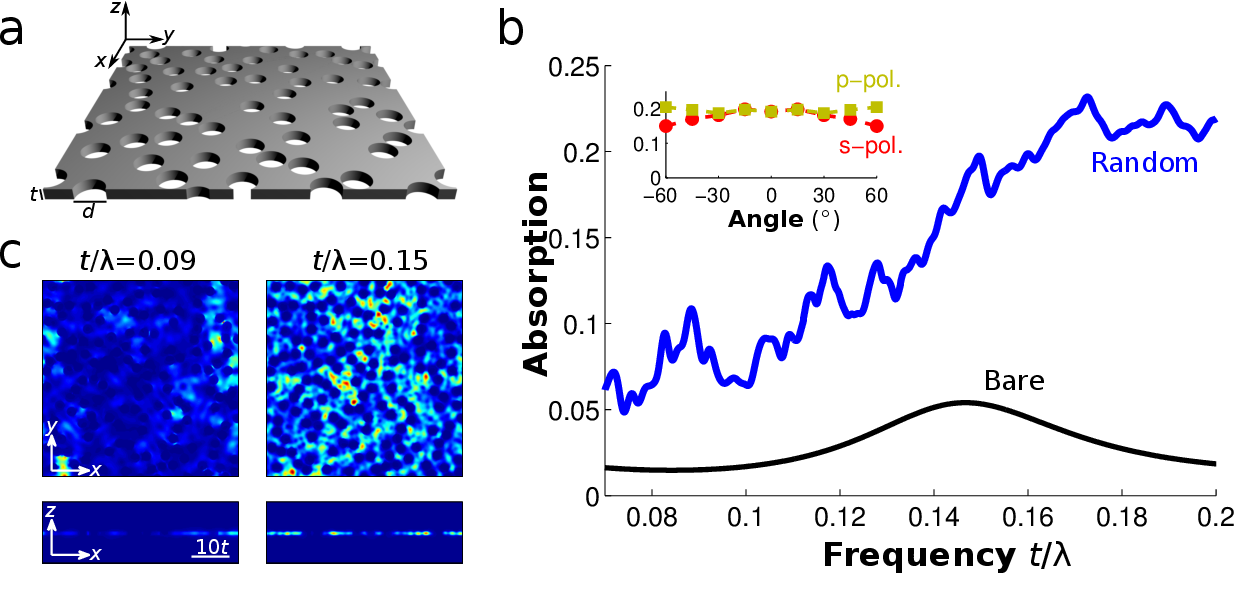}
\end{center}	
\caption{\textbf{Light trapping in thin random films.} (a) Sketch view of the randomly-nanostructured film. (b) Absorption spectra of the bare and patterned films (black and blue lines, respectively) at normal incidence. Inset: angular dependence of the absorption at $t/\lambda=0.15$ for $s$- and $p$-polarized incident light (red circles and yellow squares, respectively). The random pattern of holes leads to a large absorption of the incident light over a wide angular range and for both polarizations. (c) Top and side views of the electromagnetic energy density inside a non-absorbing sample for a normally-incident planewave at frequencies $t/\lambda=0.09$ (left) and $t/\lambda=0.15$ (right). The normalization on the color scale is made according to the highest value at the specific frequency. The high energy density in the film implies an efficient light trapping, which explains the high absorption efficiency. The speckle pattern fluctuations differ at these frequencies, indicating different transport regimes (see the Supplementary Information).
\label{fig1}}
\end{figure*}

Interestingly, only little attention has been given so far to \textit{realistic} two-dimensional disordered structures for which the third dimension is intrinsically part of the physics involved. An example of such a case is that of a thin dielectric membrane containing a random pattern of circular holes going through the entire film thickness~\cite{Riboli2011, Noh2011}, as illustrated in Fig.~\ref{fig1}a. Light is confined to the plane of the film by index guiding and due to the presence of the holes, which serve as scattering centers, the light waves are multiply-scattered in the plane, giving rise to a transport that is essentially two-dimensional. Because of the finite thickness of the film, however, the electromagnetic modes of the system are leaky, and thus, can be optically excited from the third dimension. Since these modes intrinsically depend on the optical transport properties of the system, the question we will address in this Letter is whether such two-dimensional disordered structures can be used to \textit{control} the light coupling process between a thin film and free space.

Gaining control over such a process through the optical modes of two-dimensional disordered structures constitutes a novel approach for photon management. Thin-film technologies have played an increasingly important role in energy efficiency, constituting an excellent opportunity for efficient and scalable solar cells~\cite{Brown2009}, as they benefit from a lower amount of material used and a better collection of the charges, and high-intensity lighting devices. Techniques involving random patterns, such as textured surfaces~\cite{Schnitzer1993, Wiesmann2009, Rockstuhl2010, Ferry2011, Sheng2011} -- initially applied to thick layers~\cite{Yablonovitch1982,Green2002} -- take advantage of the broadband light spreading they provide to improve the light coupling efficiency between free space and the films. In contrast, we propose a novel scheme in which the nanopatterning gives rise to disordered quasi-guided modes formed by an engineered two-dimensional multiple-scattering process.

To start, we demonstrate that the introduction of a random pattern of holes, as in Fig.~\ref{fig1}a, in a thin absorbing film can lead to a large and broadband absorption enhancement. For this, we calculate the absorption spectrum for a planewave in air at normal incidence on the nanostructured film and compare it with that of the bare film. The numerical analysis was performed by the three-dimensional Finite-Difference Time-Domain method~\cite{Taflove2005} using periodic boundary conditions on a large area to mimic infinitely large structures. For convenience, all dimensions in this study are given as a function of the film thickness $t$ and can be rescaled according to one's need. Here, the circular holes have a diameter $d=2.7t$ and filling fraction $f=30\%$. The material of the film has a permittivity $\varepsilon=12$ and an absorption length $\ell_i=33t$, which, for the sake of the analysis, was considered constant over the wavelength range of interest. This corresponds approximately to the absorption of a thin film ($t \approx 100$ nm) of amorphous Silicon (a-Si) in the red part of the spectrum~\cite{Palik1997}. Note however that the concepts presented below are valid for any type of two-dimensional random systems of finite thickness and can be applied to any inorganic or organic material.

The absorption of the patterned film is plotted in Fig.~\ref{fig1}b together with the absorption of the bare film. We find that the random pattern of holes enhances significantly the absorption of the film over the entire wavelength range under consideration, the integrated absorption being increased by a factor $F=4.8$. This absorption enhancement occurs in spite of the fact that $30\%$ of absorbing material has been removed. The inset of Fig.~\ref{fig1}b shows the angular dependence of the absorption at $t/\lambda=0.15$ for both $s$- and $p$-polarizations. One can see that the absorption remains particularly high even at large angles and for both polarizations of the incident light. This is a property of paramount importance for solar panels since it allows to retain a high efficiency also for diffuse illumination and at all times of the day.

To gain insight into the mechanism behind this strong absorption enhancement, it is instructive to observe how the electromagnetic energy density is distributed in the nanostructured film. The repartition of the energy density between a dielectric material and free space allows to determine the maximal enhancement of absorption in the material~\cite{Yu2010, Callahan2012}. In Fig.~\ref{fig1}c, the steady-state electromagnetic energy density distribution in the film is shown for two different frequencies ($t/\lambda=0.09$ and $0.15$), considering again an incoming continuous planewave polarized along $x$ at normal incidence on a non-absorbing patterned film with the same structural parameters as in Fig.~\ref{fig1}b. The remarkably high energy density in the film compared to free space (enhancement factors of about 16 and 40 at $t/\lambda=0.09$ and $0.15$, respectively) indicates a very efficient light trapping. The superposition of the externally excited in-plane modes yields a complex speckle pattern characterized by fluctuations of varying amplitude as a function of frequency. Speckle intensity fluctuations in random media derive from wave interferences between the multiple scattering paths and are known to carry important information about transport~\cite{Akkermans2007}. In two-dimensional systems of finite thickness, they are expected to rely on a delicate interplay between in-plane multiple scattering, localization phenomena, and out-of-plane leakage (for further study, see the Supplementary Information).

We now investigate the coupling process between free space and the modes of the two-dimensional random medium. Information about the coherent transport of waves in complex systems is given by the so-called spectral function~\cite{Sheng2010,Lagendijk1996}, which essentially counts the number of modes that can be excited in the system as a function of frequency and wavevector. Upon integration over wavevectors, the spectral function equals the density of states per unit volume of the system. In the present case, the repartition of such resonances in reciprocal space can simply be accessed through an analysis of the speckle field distributions induced by the in-plane multiple light scattering. Indeed, the complex field pattern in real space is due to the superposition of the different optical modes in the medium and reflects the underlying transport process. All these modes can be separated by a Fourier transform of the fields, each point in reciprocal space corresponding to a certain number of resonances~\cite{Engelen2007, Burresi2011}. The resulting resonance density distribution therefore provides a map of all the possible excitations to the optical modes of the system.

Since in our case the transport is essentially in the plane of the film, this Fourier analysis of the field can be made in purely two-dimensional systems. For the sake of the demonstration, we focus on the lowest-order TE-polarized propagating mode of the slab that is the most relevant for such thicknesses. The refractive index of the dielectric material is set to the effective (frequency-dependent) refractive index of the propagating mode as to account effectively for the finite thickness of the film. A large number of $H_z$ dipole sources with random phases are placed at random positions in the computational domain to excite all possible modes and the spatial field maps are recorded after a long time.

\begin{figure*}
\begin{center}	
		\includegraphics[width=0.8\textwidth]{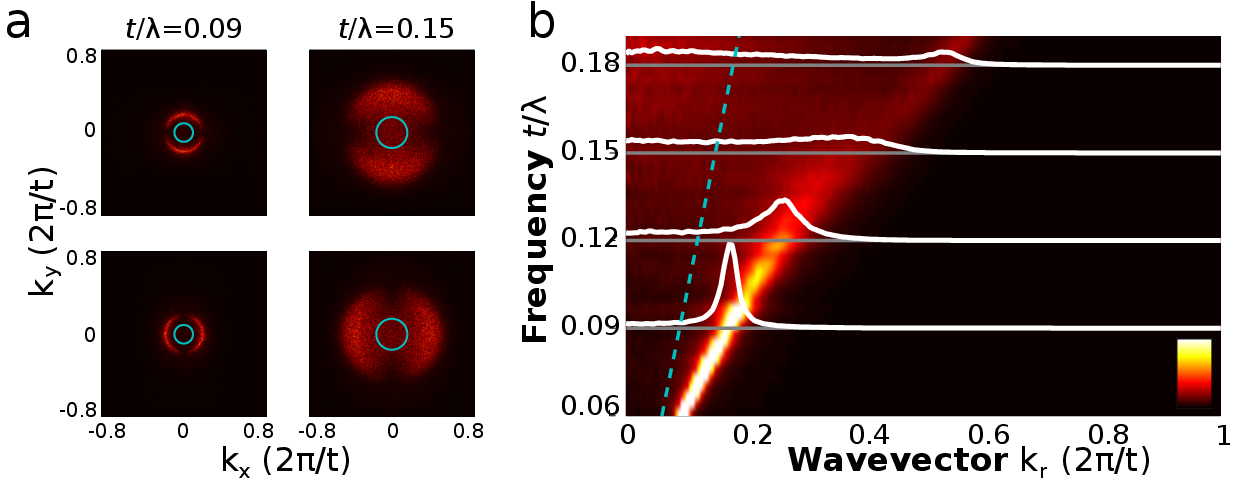}
\end{center}
\caption{\textbf{Distribution of resonances in two-dimensional random media.} (a) Norm of the Fourier transform of the electric field components in the plane of the film, $|\tilde{E}_x|$ and $|\tilde{E}_y|$, at the frequencies $t/\lambda=0.09$ (left) and $t/\lambda=0.15$ (right). Light incident on such a random medium from the third dimension can couple to the structure by the optical excitation of the resonances lying in the radiative zone (central disk delimited by a light blue solid circle). (b) Radial-averaged density of resonances $M(t/\lambda,k_r)$ in the two-dimensional random structure. The white lines are cross-cuts of the resonance distribution at frequencies $t/\lambda=0.09$, $0.12$, $0.15$ and $0.18$. The light line is shown as a light blue dashed line. The well-defined peak at lower frequencies, which corresponds to a propagating mode with a certain effective refractive index, broadens at higher frequencies due to the stronger light scattering by the holes, which in turn modifies the distribution of leaky resonances.
\label{fig2}}
\end{figure*}

The norm of the Fourier components of the electric field along the $x$- and $y$-directions in the system are shown in Fig.~\ref{fig2}a at frequencies $t/\lambda=0.09$ and $t/\lambda=0.15$. The two lobes observed for both field components and at both frequencies correspond to the most probable directions of propagation and their associated effective propagation constant. The coupling process between in-plane modes and free space, of particular importance in this study, translates into a phase-matching condition that is met only for in-plane wavevectors $\textbf{k}_{||}$ lying within the so-called radiation zone, i.e. for $|\textbf{k}_{||}|<k_0$, with $k_0=2\pi/\lambda$ the wavevector in free space ($|\textbf{k}_{||}|=0$ for normal incidence). By making use of the isotropy of the structure, we calculate the radial-averaged density of resonances $M(t/\lambda,k_r)= |\tilde{E}_x(t/\lambda,k_r)|^2+|\tilde{E}_y(t/\lambda,k_r)|^2$ with $k_r=|\mathbf{k}_{||}|$ at different frequencies (see Fig.~\ref{fig2}b) to describe the dispersion of the average propagating mode in the random medium. At lower frequencies, the resonance distribution exhibits a well-defined peak, corresponding to a propagating mode with an effective refractive index given by the Maxwell-Garnett mixing rule (see the Methods section). The progressive broadening of the peak in reciprocal space with increasing frequency is associated with a stronger light scattering by the holes. When the wavelength in the medium becomes comparable to the length scale over which the refractive index fluctuates (at $t/\lambda \approx 0.135$), the resonances broadly spread over the reciprocal space, leading to an increase of the resonance density in the radiative zone and thus, to a favoured optical excitation. The almost uniform distribution of the resonances over the radiative zone also explains why the coupling efficiency depends only weakly on the angle of incidence of light from the third dimension.

In sum, this purely two-dimensional analysis made it possible to grasp the essential physics of the light coupling process between free space and a thin nanopatterned film, showing that the density of leaky resonances is directly related to the multiple light scattering in the plane of the film through the formation of disordered optical modes. As we will now show, it is possible, in addition, to optimize the structural parameters of a disordered structure to tune the efficiency of the optical excitation process and calibrate its spectral response to a desired wavelength band.

We have considered so far a random structure in which the position of the holes was uncorrelated (apart from a non-overlap constraint) and have seen that its optical response does not exhibit sharp spectral and angular features. These optical characteristics, however, can be changed by adding structural correlations in the disordered pattern as to provide an additional degree of freedom to control light transport~\cite{RojasOchoa2004}. In two-dimensional structures, correlations have been used to create large and isotropic photonic band gaps~\cite{Florescu2009} as well as to gain a better control over random lasing processes~\cite{Noh2011}. To illustrate the possibilities that structural correlations offer for light coupling purposes, we investigate the optical properties of an \textit{amorphous} structure that possesses a statistical short-range correlation in the position of the holes but lacks any long-range order (e.g. periodicity). Note that the pattern of the amorphous structure does not contain large ordered domains (see Methods).

\begin{figure*}
\begin{center}	
		\includegraphics[width=1.0\textwidth]{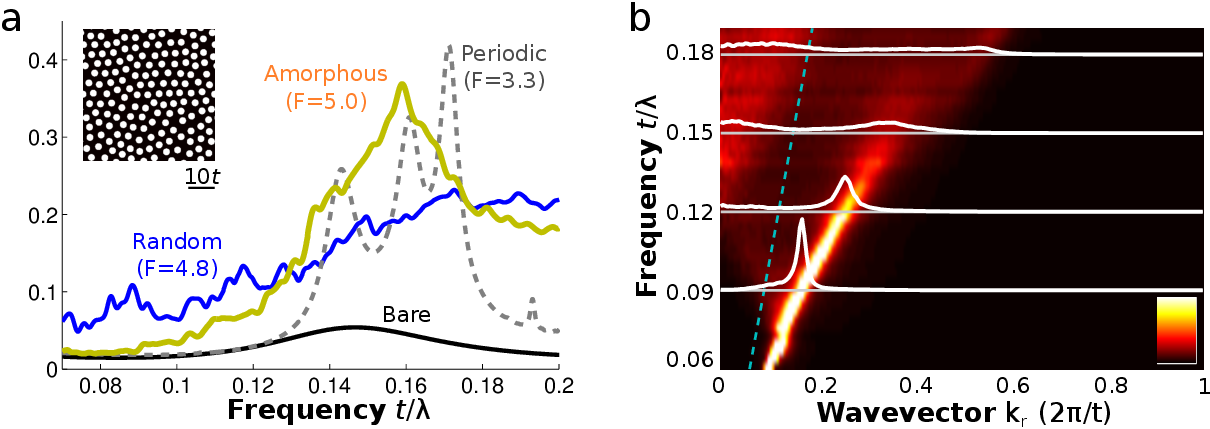}
\end{center}	
\caption{\textbf{Influence of structural correlations.} (a) Absorption spectrum of the two-dimensional amorphous structure (orange solid line), shown in the inset, compared with the absorption spectrum of the bare film (black solid line), the random structure (blue solid line) and the periodic structure (gray dashed line). (b) Radial-averaged density of resonances $M(t/\lambda,k_r)$, as in Fig.~\ref{fig2}b, for the two-dimensional amorphous structure. The amorphous structure, due to structural correlations, leads to a larger increase of the absorption on a smaller spectral range compared to the simple random case. The reduced and enhanced absorption are explained by the depletion and increase of the density of resonances for a planewave at normal incidence on the film at lower and higher frequencies, respectively.
\label{fig3}}
\end{figure*}

We calculate the absorption spectrum of a thin film containing a two-dimensional amorphous pattern of holes with the same diameter and at the same filling fraction, see Fig.~\ref{fig3}a. The effect of short-range correlations on the absorption is striking. While the simple random structure provides a broadband absorption enhancement over the entire range of interest, the amorphous structure exhibits an absorption comparable to that of the bare film at lower frequencies, followed by a large absorption enhancement at frequencies $t/\lambda \approx 0.16$, for an integrated absorption enhancement factor $F=5.0$ with respect to the bare slab.

This behavior is perfectly explained by the distribution of resonances shown in Fig.~\ref{fig3}b for the amorphous case. At frequencies below to $t/\lambda \approx 0.07$, a depletion of the density of resonances in the radiative zone is observed. Due to their very low number, light incident from free space on the film crosses it almost without coupling to the in-plane modes. This peculiar phenomenon is specific to our system of finite thickness and illustrates nicely that the coupling process (in both the random and amorphous cases) is not a property of the single scatterer but that of a collection of neighbouring scatterers.

This depletion of leaky resonances is in fact accompanied by two other remarkable effects, both of which are inherent to the short-range correlation in the structure and evidently do not require a long-range periodicity. First, at a frequency $t/\lambda \approx 0.07$, corresponding to the Bragg condition for which the wavelength in the medium $\lambda_e=\lambda/n_e$ equals twice the typical distance $a$ between nearest neighbours, $\lambda_e=2a$, the dispersion undergoes a strong deformation. The lattice resonance is expected to enhance the in-plane scattering strength in this range~\cite{Yang2010}. Second, above this frequency, a branch with a \textit{negative} slope is observed, albeit in a disordered system, strongly reminding the dispersion relations of periodically-nanostructured photonic crystals~\cite{Joannopoulos2008}. On the occurance of this new dispersion branch for $k_r=0$, the number of leaky resonances is strongly increased, explaining the stronger absorption in this range. Furthermore, while the coupling to the random system is expected to be efficient at all angles, the folded branch in the amorphous structure makes such that coupling at an angle may be more efficient in some frequency range. Disorder correlations therefore allow for a tuning of both the spectral and the angular light coupling of two-dimensional disordered media. For applications in lighting, one could take advantage of the negatively-sloped branch in amorphous structures to design light-emitting diodes with some \textit{iridescent diffuse} light. All in all, structures with correlated disorder are promising candidates for future integrated optical devices, as imposing only a weak constraint on the position of the scattering elements, such as a minimum distance, is enough to induce radically different optical characteristics.

It is also interesting to compare the performance of the amorphous structure with that of the corresponding periodic structure. To this aim, we consider a hexagonal lattice with the same holes and same filling fraction as the disordered structures. The absorption spectrum, shown in Fig.~\ref{fig3}a, exhibits three peaks in correspondance with three Bloch modes of the photonic crystal structure accessible at normal incidence, for an integrated absorption enhancement factor $F=3.3$ with respect to the bare slab. Since the typical distance between neighboring holes is comparable in the periodic and amorphous structures, the high absorption frequency range arising due to the short-range order comes at a nearby frequency. It is important to note that the absorption spectrum of the amorphous structure is \textit{not} a mere broadening of that of the periodic structure. The absorption efficiency of the amorphous structure is found to be comparable to that of the periodic structure on resonance and covers a broader spectral range. This suggests that disordered structures could actually be more efficient than periodic ones depending on the frequency range and bandwidth of interest.

To support this latter statement and motivate further investigation into the use of two-dimensional disordered structures for solar energy, we calculate the integrated absorption efficiency of nanopatterned thin a-Si films in the 600-800 nm wavelength range of the solar spectrum, of particular relevance for photovoltaics. At smaller wavelengths, when the absorption length of the material becomes smaller than the sample thickness, single scattering by the holes is expected to be sufficient to compensate for the material removal~\cite{Meng2011}, suggesting also that different hole patterns should yield comparable efficiencies. The films have a thickness $t=100$ nm, which is comparable with the diffusion length of the minority charge carriers in a-Si, and contain air holes of diameter $d=350$ nm at a filling fraction $f=30\%$ distributed according to the patterns considered above. Note that these structural parameters are close to the optimal ones for light trapping by photonic crystal structures in thin films in the visible range~\cite{Meng2011}.

The results are shown in Table~\ref{tab:absorption_efficiency} (see the spectra in the Supplementary Information). Interestingly, the random pattern is found to be as efficient as the periodic one in this range with an integrated absorption of about $53\%$, that represents an absorption enhancement factor $F=3.3$ with respect to the unpatterned film. As for the film containing the amorphous pattern, it exhibits a $10\%$ higher absorption efficiency compared to that of the previous structures. Such a noticeable enhancement results from the combined effect of an improved light trapping due to short-range correlations and of a broad operation bandwidth due to its disordered nature.

\begin{table}
\centering
\begin{tabular}{ | c | c |}
  \hline
  \textbf{Thin a-Si film} & \textbf{Integrated absorption efficiency} \\
  \hline \hline
  Bare (unpatterned) & $16\%$ \\
  \hline
  Periodic pattern & $53\%$ \\
  \hline
  Random pattern & $53\%$ \\
  \hline
  Amorphous pattern & $58\%$ \\
  \hline
\end{tabular}
\caption{\textbf{Thin a-Si film.} Integrated absorption efficiency of thin films of a-Si of thickness $t=100$ nm in the wavelength range $600 \leq \lambda \leq 800$ nm (see the spectra in the Supplementary Information). The nanopatterned films contain air holes of diameter $d=350$ nm with a filling fraction $f=30\%$. While the periodic and random structures exhibit similar efficiencies, in spite of the fact that the periodic structure is already very efficient, the amorphous structure is more efficient by about $10\%$ on this wavelength range. No structural optimization has been made on the disordered structures.}
\label{tab:absorption_efficiency}
\end{table}

Towards applications in photovoltaics, let us finally note that the model of Lambertian light trapping for such a film including a backreflector~\cite{Green2002}, albeit derived in the ray-optics limit, predicts a \textit{limiting} efficiency of approximately $93\%$ for diffuse incident light. Considering that no optimization has been made so far on the structural parameters of the disordered structures and that high absorption efficiencies are also expected at oblique incidence, we are confident that the performance of photonic structures with engineered disorder in solar cell configurations could be improved greatly.

Our finding that two-dimensional disordered structures make it possible to control the light trapping and absorption in thin films on broad spectral and angular ranges provides a powerful approach for photon management in energy efficiency technologies and may form a new generation of high-efficiency thin-film photovoltaic devices and light sources. The observation that disordered structures can have comparable and possibly higher efficiencies compared to their periodic counterpart also constitutes an interesting aspect on the technological level, as releasing the constraint of periodicity could lead to the elaboration of more efficient and cost-effective routes for large-scale fabrication. These results should encourage deeper investigations onto the fabrication and characterization of more specific solar-cell architectures and their comparison with alternative techniques, such as randomly-textured surfaces and periodic or nearly-periodic photonic patterned films. Our study also raises more fundamental questions such as the role of strong interference effects, such as light localization, in the coupling process. We believe that the wide knowledge of the scientific community in the physics of waves in random and complex media can push the energy efficiency research to a higher level. Finally, since the concepts presented in this Letter rely on the properties of waves in general, applications in different fields could also be conceived.

\section*{Methods}
\textbf{Structural and material parameters.} The random patterns, on one hand, were generated by Random Sequential Addition of hard circular cylinders at a filling fraction of $30\%$ on a square area with periodic boundary conditions. The amorphous patterns, on the other hand, were generated by the Lubachevsky-Stillinger algorithm on the same area~\cite{Skoge2006}. This algorithm was used to impose short-range disorder correlations in the position of the cylinders. To avoid the formation of large ordered clusters, as in polycrystalline structures, a polydispersity of $1.1$ in the size of the cylinders was used. The growth of the cylinders was stopped at a filling fraction of $69\%$ and their diameter was set to a constant diameter such that the final filling fraction would be $30\%$.

For the refractive index of the dielectric medium in the two-dimensional calculations, we used the effective index of refraction of the lowest-order TE propagating mode, calculated by an online mode solver for multilayered media~\cite{Hammer} at different wavelengths. A polynomial fit was used to model the refractive index of the dielectric film over the entire range of interest: $n_s=0.433+37.1267 \left( \frac{t}{\lambda} \right) -222.756 \left( \frac{t}{\lambda} \right)^2 + 693.63 \left( \frac{t}{\lambda} \right)^3-888.889 \left( \frac{t}{\lambda} \right)^4 $. For the effective refractive index of the two-dimensional randomly patterned array, the two-dimensional Maxwell-Garnett mixing rule can be used: $n_e=n_s \sqrt{ 1+\frac{2f\alpha}{1-f\alpha} }$, with $\alpha=\frac{n_c^2-n_s^2}{n_c^2+n_s^2}$ with $n_c=1$ the refractive index of the cylinders and $n_s$ the effective refractive index of the lowest-order TE propagating mode defined above. For the calculations of the integrated absorption efficiency of thin a-Si films, the refractive index of the material was taken from the litterature~\cite{Palik1997}.

\textbf{Numerical resolution.} Calculations were performed with the Finite-Difference Time-Domain method~\cite{Taflove2005} using a freely available software package~\cite{Oskooi2010}. The mesh resolution was set to have at least 10 grid points per wavelength in the higher refractive index material. Periodic boundary conditions in the lateral directions ($x$ and $y$) were used in all calculations. The lateral size of the computational domain was $L=200t$ in the two-dimensional calculations, $L=50t$ in the three-dimensional calculations on the films with constant absorption, and $L=5$ $\mu$m in the three-dimensional calculations on the a-Si films. The results of calculations shown in the article correspond to single disorder configurations. We have performed convergence checks on the absorption spectra and intensity statistics to make sure that the average properties of the simulated systems remained unchanged (apart from local fluctuations) when using larger samples and performing an average over disorder realizations.

The absorption $A$ was calculated by recording the reflected and transmitted fluxes, $R$ and $T$, respectively, for long times under excitation by planewave pulses at different frequencies, and calculating $A=1-T-R$. For studies on broad frequency ranges and considering the large size of the computational area, this approach was found to provide accurate results in a reasonable amount of time. For the absorption efficiency of the  thin a-Si films, the calculations were made frequency-by-frequency using the same approach, for the sake of consistency, with a 10 nm wavelength resolution for the bare and periodic structures, and 25 nm wavelength resolution for the disordered structures. The integrated efficiency was finally calculated assuming the air-mass 1.5 spectrum for the incident light.

For the studies on the two-dimensional speckle field and intensity distribution of the random and amorphous structures, a total of 2000 $H_z$ dipoles, at random positions uniformly distributed throughout the entire computational domain and with random phases, was used.

\bibliographystyle{apsrev}

\section*{Acknowledgements}
This work is supported by the Eu NoE ``Nanophotonics for Energy Efficiency'', the Italian CNR project EFOR and ENI S.p.A.. We gratefully acknowledge Pierre Barthelemy, Jacopo Bertolotti and Tomas Svensson for insightful discussions.

\section*{Author contributions}
All authors developed the concept. K.V. carried out the numerical simulations. K.V. and M.B. performed the data analysis. All authors discussed and interpreted the results. K.V. prepared the manuscript with suggestions from M.B., F.R. and D.S.W.

\section*{Additional information}
Supplementary information is available in the online version of the paper. Reprints and permissions information is available online at www.nature.com/reprints. Correspondence and requests for materials should be addressed to K.V.

\section*{Competing financial interests}
The authors declare no competing financial interests.

\end{document}